\documentclass[12p]{article}
\pdfoutput=1

\usepackage{graphicx}
\usepackage{amssymb}
\usepackage[figuresright]{rotating}

\newcommand{\be}{\begin{equation}}
\newcommand{\ee}{\end{equation}}

\begin{document}

\title{Floridian high-voltage power-grid network partitioning and
cluster optimization using simulated annealing}

\author{
\centerline{Ibrahim Abou Hamad$^{1}$, Per Arne Rikvold$^1$, Svetlana V.\ Poroseva$^2$}\\
\\
\centerline{$^1$Department of Physics, Florida State University}\\
\centerline{Tallahassee, FL 32306-4350, U.S.A.}\\
\centerline{$^2$Mechanical Engineering Department, University of New Mexico}\\
\centerline{Albuquerque, NM 87131-0001, U.S.A.}
}

\maketitle

\begin{abstract}
Many partitioning methods may be used to partition a network into smaller 
clusters while minimizing the number of cuts needed. However, other
considerations must also be taken into account when a network represents
a real system such as a power grid. In this paper we use a simulated annealing
Monte Carlo (MC) method to optimize initial clusters 
on the Florida high-voltage power-grid network that were formed by associating
each load with its ``closest'' generator. The clusters are optimized 
to maximize internal connectivity within the individual clusters and 
minimize the power deficiency or surplus that clusters may otherwise have.
\end{abstract}



\section{Introduction}
\label{sec-Int}
The lack of a pre-planned 
strategy for splitting a power-grid system into separate parts with
self-sufficient power generation is one of the reasons for large-scale blackouts
that have devastating effects on the economy and welfare of any modern
society~\cite{LILI05,PEIR09}. This defensive strategy (intentional islanding)
is effective in preventing cascading outages~\cite{PEIR09,YANG06}.

Multiple approaches to intentional islanding
(see, e.g., \cite{LILI05,PEIR09,YANG06,KARY98,CHOW08,WANG04,LIU06}) 
have been suggested for optimizing the selection of the lines to be cut.
These studies can be based on the analysis 
of the system topology based on a representation of the network 
as a graph \cite{SEAR03,FORT10,NEWM04C,HAMAD2010}. 
Some topologies are easier to split into islands than others.
The identification of ``weak'' links and their removal can split 
a given topology into independent islands. 
While many of the above approaches are very good at the identification
of ``weak'' links, the resulting clusters or islands are usually not
optimized for other qualities such as generating capacity.

Here we present a study utilizing a matrix method for intentional
islanding of a utility power grid. The method uses a
Monte Carlo (MC) simulated annealing~\cite{SIMUL83} technique for optimizing
the resulting islands' internal connectivity as well as balancing their
generating capacity. The concept is illustrated by application to the Floridian
high-voltage power grid. 

\section{Methods}
\label{sec:met}
The quality of a particular partitioning of a graph into $M$ communities,
${\mathcal C} = \{C_1, ..., C_M\}$ can be estimated by Newman's
{\it modularity\/} \cite{NEWM04C}. It compares the proportion of edges that are
internal to a community in the particular graph with the same proportion in an
average null-model. It is defined as follows:
\be
Q = \frac{1}{w} \sum_{ij} \left( w_{ij} - \frac{w_i w_j}{w}  \right)
\delta \left( C(i),C(j) \right)
\;,
\label{eq:Q}
\ee
where $\delta \left( C(i),C(j) \right) = 1$ if nodes $i$ and $j$ belong to the
same community, and $0$ otherwise. Ideally, one would like to maximize $Q$ while
partitioning a power-grid network. This will ensure that the different
communities are well connected internally. Moreover, one would like to minimize 
the generating power surplus or deficiency over all the clusters. Here we will
use a partitioning scheme consistent with a power-grid network and try to
optimize the resulting clusters for internal connectivity and power
self-sufficiency using a Monte Carlo simulated annealing approach. The resulting
set of clusters form a new network (in a renormalization-group sense) where each
cluster is represented by a node on the new network. The islanding procedure and
MC optimization are repeated until some required criteria are met.
\subsection{Partitioning}
We use a simplified representation of the 
power grid as an undirected graph \cite{SEAR03,NEWM04C} defined by the 
$N \times N$ symmetric {\it conductivity matrix\/} $\bf W$, whose
elements $w_{ij} \ge 0$ represent the ``conductivities'' of 
the edges (transmission lines) between vertices (generators or loads)
$i$ and $j$,
\begin{equation}
\label{eq:conductivities}
w_{ij}=\frac{ {\rm number\, of\, lines\, between\, vertices\,}i\,{\rm and}\,j}{\rm{\, normalized
\, geographical \, distance}}, 
\end{equation}
where the ``geographical distance'' is the length of the edge connecting nodes
$i$ and $j$, and is normalized by the minimum geographical distance between two 
nodes over the whole network.

The row sums of $\bf W$ define the diagonal matrix $\bf D$. Graph analysis can
be performed using one of several matrices derived from $\bf W$. The Laplacian,
${\bf L} = {\bf D} - {\bf W}$, is a symmetric matrix with vanishing row sums. It
embodies Kirchhoff's laws and thus represents a simple resistor network with
conductances $w_{ij}$. Multiplied by a column vector $| \phi \rangle$ of vertex 
potentials, it yields the vector of currents entering the circuit at each
vertex, ${\bf L} | \phi \rangle = | I \rangle$. This equation can be rewritten
as ${\bf L}^{(-1)} | I \rangle = | \phi \rangle$, where ${\bf L}^{(-1)}$ is
the pseudo-inverse of the Laplacian. In other words, given a current vector,
defined as being positive at generator nodes and negative at load nodes, one can
calculate the potential vector $ | \phi \rangle $. Using this potential vector
and the matrix $\bf W$, we calculate the network-current matrix $\bf K$, whose
elements are the currents between the corresponding nodes:
${k}_{ij}=(\phi_j - \phi_i ){w}_{ij}$. Additionally using the matrix
${\bf L}^{(-1)}$, one can calculate an effective distance or equivalent
resistance ${\bf R}$ between any two nodes~\cite{KLEIN93}:
$({\bf R})_{ij}=({\bf L}^{(-1)})_{ii}+({\bf L}^{(-1)})_{jj}-2({\bf L}^{(-1)})_{ij}$.
Consequently, given a current vector and a conductivity matrix ${\bf W}$, the
network-current matrix ${\bf K}$ and the equivalent resistance matrix ${\bf R}$ 
can be evaluated. In this paper we use these two last matrices to achieve an
initial partitioning of the Floridian high-voltage grid.

\begin{figure}
\begin{center}
\includegraphics[angle=-90,width=0.48\textwidth]{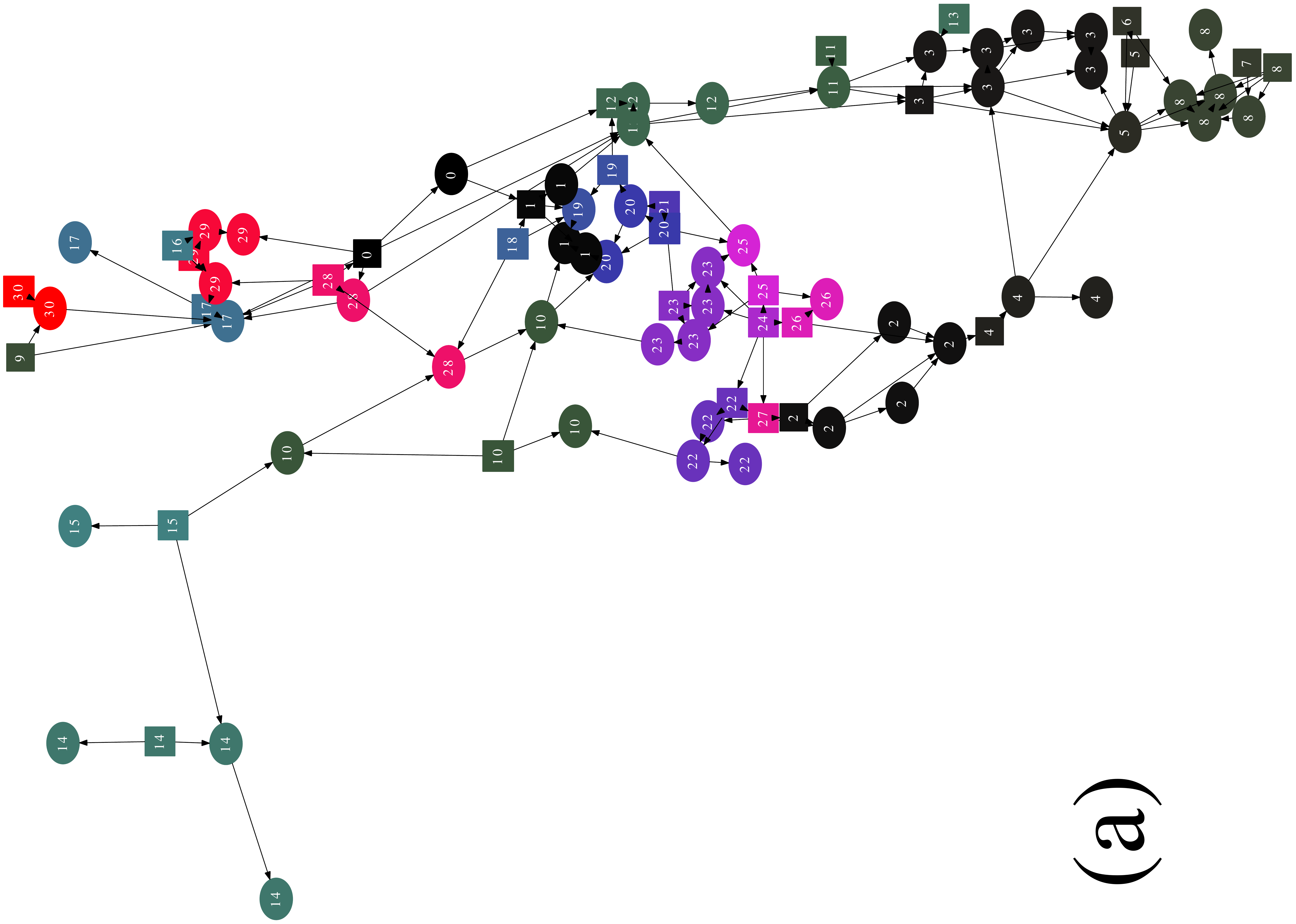} 
\hspace{0.1truecm}
\includegraphics[angle=-90,width=0.48\textwidth]{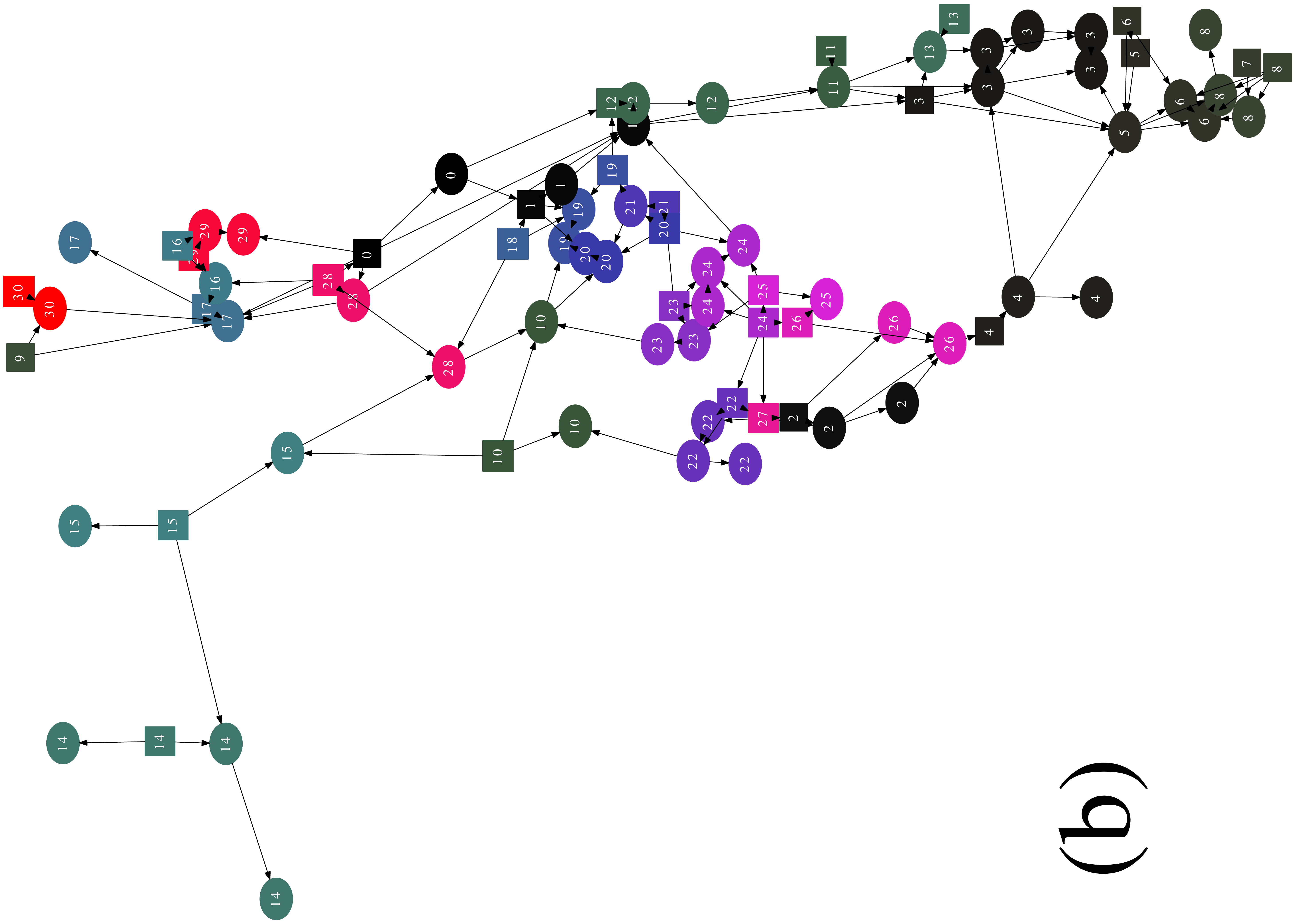} 
\end{center}
\caption[]{
(a) Florida high-voltage power grid partitioned by connecting loads (ovals) to 
their ``nearest'' generators (squares). (b) Grid clustering after the MC
simulated annealing algorithm is used for optimizing modularity and power
sufficiency of the partitioned network shown in (a). Nodes of the same color and
label belong to the same cluster. See text for discussion.
}
\label{fig:FL}
\end{figure}

The goal is to partition the power grid into communities of vertices that are
highly connected internally, but only sparsely connected to the rest 
of the network. For the islanding to be useful, each island should contain at
least one generating plant. To accomplish this, we use a clustering algorithm
where each load $i$ is connected to the ``nearest'' generator $j$. The nearest
generator $j$ is the one located ``upstream'' from load $i$, i.e.
$\phi_j > \phi_i$, and for which $({\bf R})_{ij}$ is minimum. The Floridian
high-voltage grid~\cite{FLAMAP} at this first level of islanding is shown in
figure~\ref{fig:FL}(a).

Kirchhoff's junction law, when applied at each node, tells us that the sum of all
network currents going in and out of node $i$ is equal to $ |I \rangle_i$.
Thus we can think of $ |I \rangle_i$ as the current provided by a generator
at node $i$ if it is positive, or consumed by a load if it is negative.
Moreover, given the constant voltage rating (138, 230, 345, ... kV),
$ |I \rangle_i$ becomes proportional to the power being generated or consumed at
node $i$. This means that we can choose our initial current vector proportional 
to the generating power of each power plant. Since the actual power rating for
power plants on the grid was not available to us, we here assume that each
generator's power is proportional to the number of edges linking to it, or its
{\it degree}. The current vector component at each node $i$ is then defined as
$|I \rangle_i = \frac{{\rm degree}_ i}{\sum_{\rm generators}{\rm degree}_j}\,
{\rm for\, generators}$ and
$|I \rangle_i =-\frac{{\rm degree}_ i}{\sum_{\rm loads}{\rm degree}_j}\,
{\rm for\,loads}$.

\subsection{Monte Carlo}
Since the power generation or consumption rate of any generator or load is
directly proportional to its current-vector component $ | I \rangle_i$,
a community's total generating surplus or deficiency is proportional to
the sum of its members' current-vector components. Thus, to optimize our
partitioning for well-balanced communities, we try to minimize the
variance of the new current vector $ | \tilde{I} \rangle $, whose components
are defined as $ | \tilde{I} \rangle_i = \sum_{i\ge j} |I \rangle_j \delta
(C(i),C(j))$ after each iteration of the islanding procedure.
We need to maximize the modularity $Q$ for better
internal connectivity at the same time. For this purpose, we define an
optimization parameter
\be
E=\frac{Q}{Q_{\rm init}} - \sqrt{ \frac{VAR(| \tilde{I} \rangle )}{VAR(| \tilde{I} \rangle_{\rm init})}},
\ee
where the subscript ``init'' designates the initial value {\it after}
recombination, but {\it before} any MC steps. This form gives equal emphasis on 
optimizing modularity and load balance. More emphasis could be given to the
optimization of one quantity versus the other by multiplying the term
corresponding to it by some weighting factor.

The MC process proceeds as follows. First, a load node $i$ is selected at
random. Then, if $i$ is at the edge of the cluster it belongs to, i.e,
if it is connected to a neighboring cluster, we randomly select one of the
neighboring clusters connected to $i$ and attempt to move this load to that
neighboring cluster. If this move does not break the first cluster into two
disconnected parts, the move is accepted with a Metropolis
acceptance rate~\cite{METROP00} $R=\min(1,e^{(- \beta \Delta E)})$,
where $\Delta E = E_a - E_i$ is the difference between the attempted state
and the initial state for that move, and $\beta$ is an ``inverse temperature.''
In a fashion similar to simulated annealing, we start at
a high temperature and gradually decrease it to zero while saving the
configuration for which $E$ is maximum. This process is repeated to look
for the global maximum of $E$.

\subsection{Recombination}
\begin{figure}
\begin{center}
\includegraphics[angle=-90,width=0.9\textwidth]{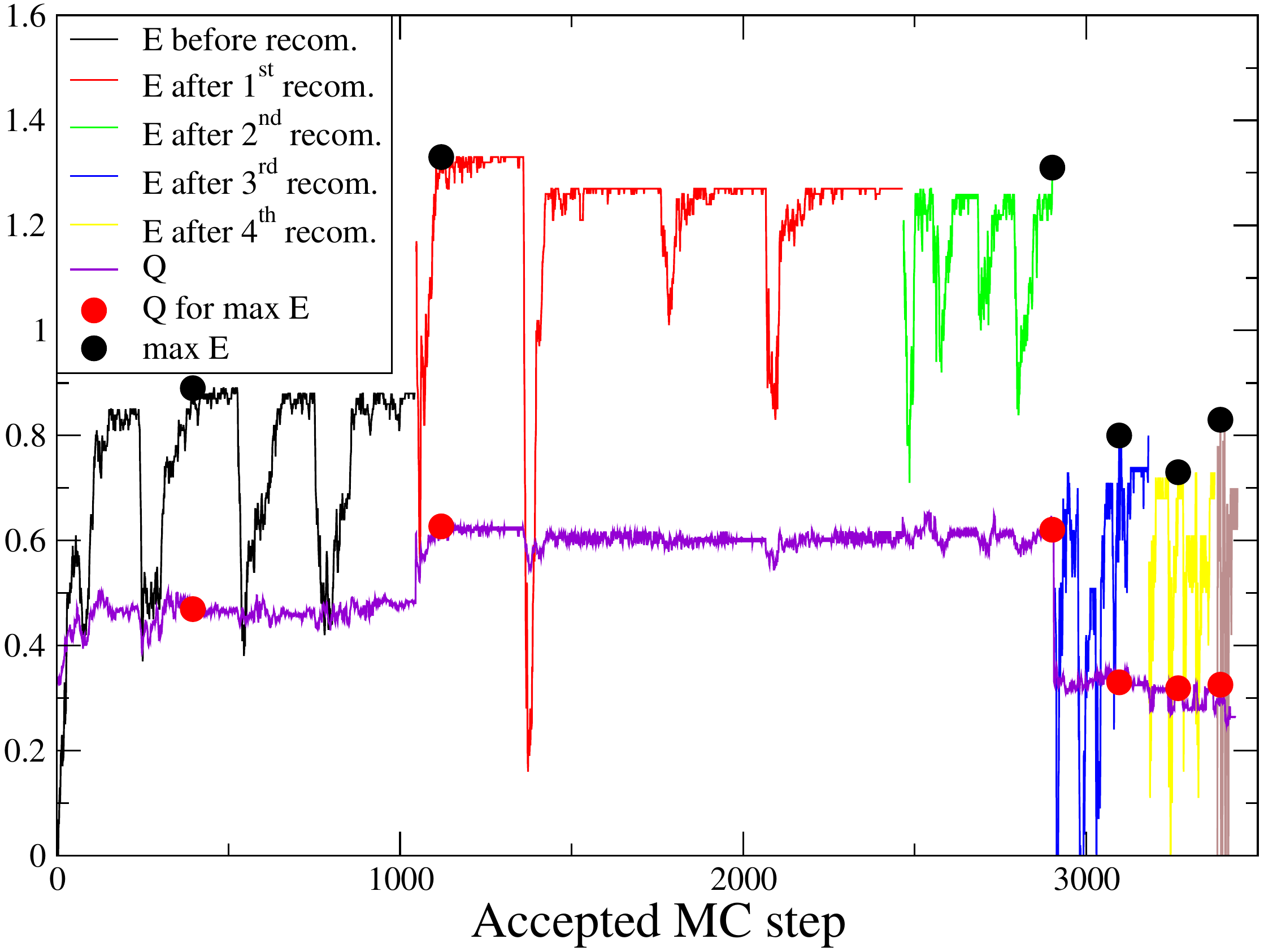} 
\caption[]{
Modularity Q and the optimization parameter $E$ ${\it vs.}$ 
MC step. 
}
\label{fig:EandQ}
\end{center}
\end{figure}

After the initial partitioning and MC, the number of clusters produced is equal
to the number of the generators in the circuit (Fig.~\ref{fig:FL}(a)), as
expected from our clustering scheme. A new network can be constructed from this
group of clusters by regarding each cluster as a new node. The connections
between the new nodes are the same as the connections between the previous
clusters. This defines a new network with new connections and a new conductivity
matrix. The current vector defined above, $|  \tilde{I} \rangle $, is the new
current vector because its components represent the generating surplus or
deficiency of each of the old clusters or new ``super-generators'' or
``super-loads,'' respectively. Given the new network and the new current vector,
we repeat the above partitioning and MC schemes on the new network. The number
of clusters at this stage is equal to the number of ``super-generators.'' This
process of recombination is repeated to look for the optimum configuration
until all the original nodes belong to one cluster. The optimization parameter
${\it vs.}$ MC step and the corresponding modularity are shown in
figure~\ref{fig:EandQ}, where the red circles are the values of $Q$ for maximum
$E$ at each level of recombination.

\section{Results and Conclusion}
\label{sec:res}
The map of the Floridian high-voltage grid \cite{FLAMAP} is a 
network with 84 vertices, 31 of which are generating plants. 
We have modeled it as an undirected graph with 
137 edges. The conductivities were calculated according to
equation~(\ref{eq:conductivities}).

\begin{figure}
\begin{center}
\includegraphics[angle=-90,width=0.9\textwidth]{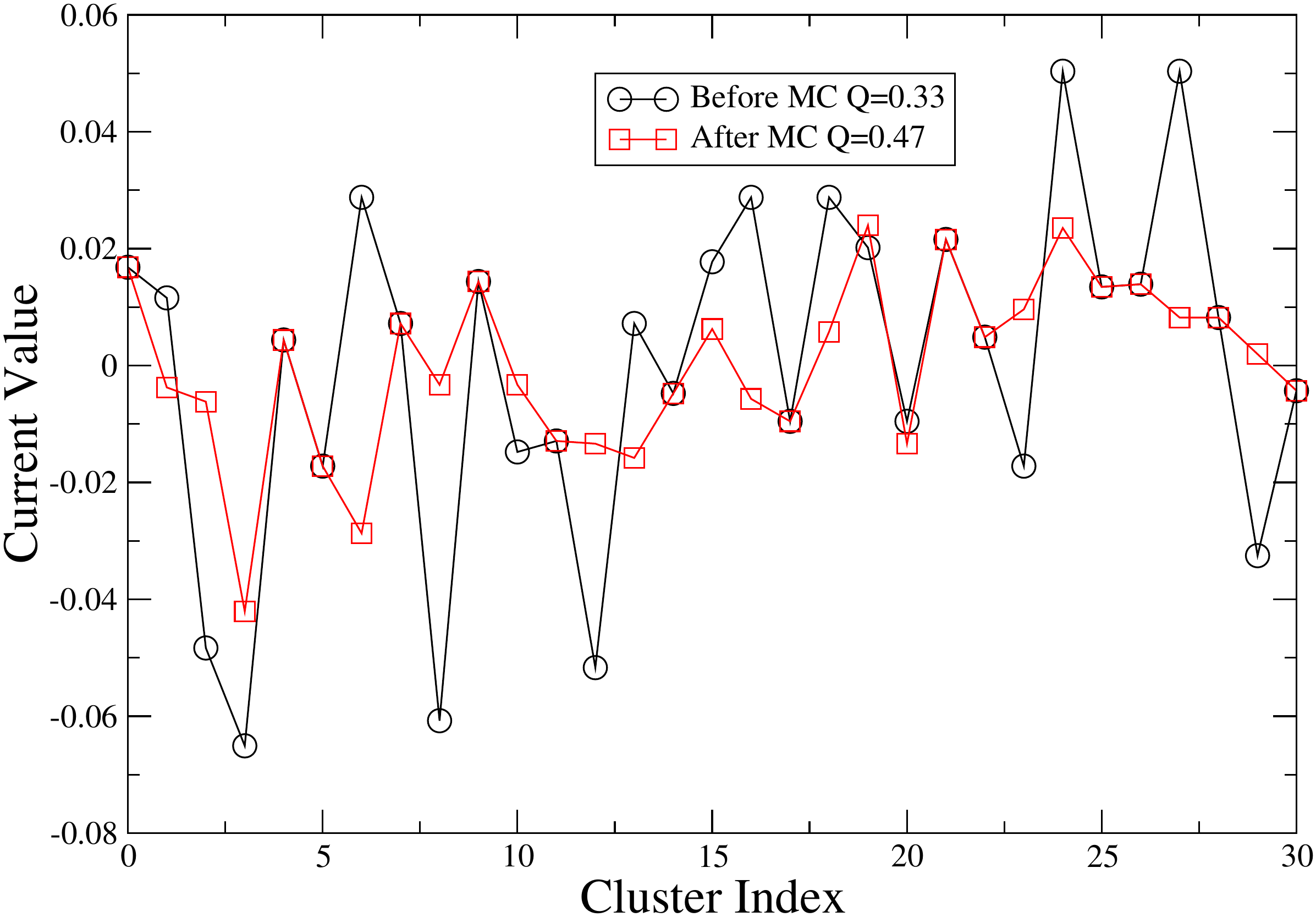} 
\caption[]{Components of the current vector $|  \tilde{I} \rangle$ {\it vs.} the
cluster index. It can be seen that the MC procedure narrows the spread of the
current values, meaning that, on average, the individual clusters are closer to self-sufficiency than before MC. The legend shows the significant
improvement in modularity.}
\label{fig:CUR}
\end{center}
\end{figure}

While figure~\ref{fig:FL}(a) shows the clusters resulting from the first
partitioning scheme, figure~\ref{fig:FL}(b) shows the same network after
the MC annealing procedure is performed. The current vector and the
corresponding modularity before and after the MC annealing are shown in
figure~\ref{fig:CUR}. As can be seen, the MC process narrows the spread of
the current values or in other words, the average power surplus or deficiency
for the clusters is smaller. Moreover, while the modularity starts at a value of
$0.33$, it ends at a value of $0.47$ after MC optimization and before the first
recombination. The maximum optimization parameter, $E_{\rm max}=1.33$ with a
corresponding $Q=0.63$ was achieved shortly after the first recombination.
Comparable values ($E=1.31$ and $Q=0.62$) are obtained in the second iteration.

While many methods can be used to partition a network into smaller clusters,
there remains the need for further optimization of the resulting cluster
properties. Here we have used a clustering procedure to partition the
Floridian power-grid network that takes into account the generating power of
each of the power plants. Moreover, we have used MC simulated annealing to
optimize the resulting clusters for better internal connectivity and power
self-sufficiency.

\section*{Acknowledgments}
\label{sec:ack}
This work was supported in part by 
U.S.\ National Science Foundation Grant No.\ DMR-0802288, 
U.S.\ Office of Naval Research Grant No.\ 
N00014-08-1-0080, and the Institute for Energy Systems, Economics, 
and Sustainability at Florida State University. 




\bibliographystyle{elsarticle-num}
\bibliography{graph}



\end{document}